\documentclass[12pt]{article}
\usepackage{amsmath}
\usepackage{graphicx}

\begin{document}
\begin{center}
\section*{Strong Decays of Baryons}
\end{center}
\begin{center}
{\bf T. Melde}$^{1}$, {\bf L. Canton}$^2$, {\bf W. Plessas}$^1$, and 
{\bf R. F. Wagenbrunn}$^1$\\
\vspace{0.3cm}
$^1$Theoretical Physics, Institute for Physics, University of Graz,\\
Universit\"atsplatz 5, A-8010 Graz, Austria\\
\vspace{0.3cm}
$^2$INFN Sezione di Padova and Dipartimento di Fisica, \\
Via Marzolo 8, Padova, Italy
\end{center}
\begin{abstract}
A Poincar\'e-invariant description of mesonic baryon resonance decays 
is presented following the point form of relativistic quantum mechanics.
In this contribution we focus on pionic decay modes. It is found that 
the theoretical results in general underestimate the experimental ones
considerably. Furthermore, the problem of a nontrivial
normalization factor appearing in the definition of the decay 
operator is investigated. The present results for decay widths suggest 
a normalization factor that is consistent with the choice adopted for the 
current operator in the studies of electroweak nucleon form factors.
\end{abstract}

\vspace{0.6cm}
\begin{center}
{\bf \large Introduction}
\end{center}
\vspace{0.3cm}

Constituent quark models (CQMs) provide an effective tool to describe 
the essential hadronic properties of low-energy quantum 
chromodynamics. 
Recently, in addition to the traditional CQM, whose 
hyperfine interaction derives from one-gluon exchange (OGE)
\cite{Capstick:1986bm}, 
alternative types of CQMs have been suggested such as 
the ones based on instanton-induced (II) 
forces~\cite{Loring:2001kv,Loring:2001kx} or Goldstone-boson-exchange 
(GBE) dynamics~~\cite{Glozman:1998ag}. In particular, the GBE 
CQM aims to include the effective degrees of freedom
of low-energy QCD, as they are suggested
by the spontaneous breaking of chiral symmetry (SB$\chi$S). 

Over the years, a number of valuable insights in strong decays of
baryon resonances have been gained by various groups, e.g., in 
refs.~\cite{Stancu:1989iu,Capstick:1993th,%
Geiger:1994kr,Ackleh:1996yt,Theussl:2000sj}. 
Nonetheless, one has still not yet arrived at 
a satisfactory explanation especially of the $N$ and $\Delta$ 
resonance decays. 
This situation is rather disappointing from the theoretical side, 
especially in view of the large amount of experimental data 
accumulated over the past years~\cite{Dytman:2003pi}. 

Here, we study the mesonic decays of baryon resonances  along 
relativistic, i.e. Poincar\'e-invariant, quantum mechanics
\cite{Keister:1991sb}. This approach is a-priori distinct 
from a field-theoretic treatment. It assumes a finite number of 
degrees of freedom (particles) and
relies on a relativistically invariant mass operator with the 
interactions included according to the Bakamjian-Thomas 
construction~\cite{Bakamjian:1953} thereby ful\-filling all 
the required symmetries of special relativity.
We assume a decay operator in the point-form spectator
approximation (PFSA) with a pseudovector coupling. 
The PFSA has already been applied to the calculation of 
electromagnetic and axial form factors of the
nucleons~\cite{Wagenbrunn:2000es,Glozman:2001zc,Boffi:2001zb} and
electric radii as well as magnetic moments of all octet and decuplet 
baryon ground states~\cite{Berger:2004yi}.
In all cases the experimental data are described suprisingly well
within this approach. 

Covariant results for the strong decays of $N$ and $\Delta$ resonances 
have 
already been 
presented in ref.~\cite{Melde:2002ga} for the relativistic GBE and 
OGE CQMs. 
They show a dramatically different behaviour as compared to previous 
non-relativistic 
calculations~\cite{Krassnigg:1999ky,Plessas:1999nb}. Specifically, it 
turns out that the theoretical
results, in general, underestimate the experimental ones 
considerably. 
This behaviour has also been observed in the relativistic calculation
based on the Bethe-Salpeter 
equation using instanton-induced dynamics~\cite{Metsch:2004qk}. 
Up till now all relativistic approaches face the problem of defining 
appropriate decay 
operators. Usually one has resorted to simplified versions such as the 
spectator model.

\vspace{0.6cm}
\begin{center}
{\bf \large Theory}
\end{center}
\vspace{0.3cm}

Generally, the decay width $\Gamma$ of a resonance is defined by the
expression
\begin{equation}
\label{eq:decwidth}
	\Gamma=2\pi \rho_{f}\left| F\left(i\rightarrow
	f\right)\right|^{2},
\end{equation}
where $ F\left(i\rightarrow f\right)$ is the transition amplitude
and $\rho_{f}$ is the phase-space factor. In eq.~(\ref{eq:decwidth}) 
one has to average 
over the initial and to sum over the final spin-isospin projections. 
Previous calculations,
based on nonrelativistic approximations of the transition amplitude 
encountered an ambiguity
in the proper definition of the phase-space 
factor~\cite{Geiger:1994kr,Kumano:1988ga,Kokoski:1987is}. Here, we 
present a 
Poincar\'e-invariant definition of the transition amplitude, thereby 
resolving this ambiguity.
In particular, we adhere to the point-form of relativistic quantum 
mechanics~\cite{Keister:1991sb}, because in this case the generators 
of the
Lorentz transformations remain purely kinematic and the theory is 
manifestly 
covariant~\cite{Klink:1998pr}. The interactions are introduced into 
the (invariant) mass operator following the Bakamjian-Thomas
construction~\cite{Bakamjian:1953}.
The transition amplitude for the decays is defined in
a covariant manner, under overall momentum conservation
($P'_{\mu}-P_{\mu}=Q_{\pi,\mu}$), by
\begin{equation}
      F\left(i\rightarrow 
       f\right)=\left< P,J,\Sigma\right|
       \hat {\cal D}_{\alpha}
       \left|P',J',\Sigma'\right>\, .
\end{equation}
Here 
$\left<P,J,\Sigma\right|$ and $\left|P',J',\Sigma'\right>$ 
are the eigenstates of the decaying resonance and the nucleon ground 
state, respectively.
Inserting the appropriate identities leads to the reduced matrix 
element
\begin{eqnarray}
F\left(i\rightarrow f\right)
    &\sim & 
    \sum_{\sigma_{i},\sigma'_{i}}
\sum_{\mu_{i},\mu'_{i}
}
\int{      d^{3}k_{2}d^{3}k_{3}d^{3}k'_{2}d^{3}k'_{3}    
	     }
	     \nonumber
	     \\
	     & &
	      \Psi^{\star}_{MJ\Sigma}
  \left(
  \vec k_{1},\vec k_{2},\vec k_{3};\mu_{1},\mu_{2},\mu_{3}
  \right)
  \Psi_{M'J'\Sigma'}
  \left(
  \vec k'_{1},\vec k'_{2},\vec k'_{3};\mu'_{1},\mu'_{2},\mu'_{3}
  \right)
	     \nonumber \\
	     & &
	     \prod_{\sigma_{i}}{
	     D^{\frac{1}{2}\star}_{\sigma_{i}\mu_{i}}
	     \left[R_{W}\left(k_{i},B\left(v_{in}\right)\right)\right]}
	     \nonumber
	     \\
	     & &
	     \left<{p_{1}},{p_{2}},
	     {p_{3}};\sigma_{1},\sigma_{2},\sigma_{3}\right|
	     \hat {\cal D}_{\alpha}
	     \left|{ p'_{1}},{ p'_{2}},
	     { p'_{3}};\sigma'_{1},\sigma'_{2},\sigma'_{3}\right>
	     \nonumber\\
	     & & 
	     \prod_{\sigma'_{i}}{
	     D^{\frac{1}{2}}_{\sigma'_{i}\mu'_{i}}
	     \left[R_{W}\left(k'_{i},B\left(v_{f}\right)\right)\right]
	     }\, ,
\end{eqnarray}
where the rest-frame baryon wave functions $\Psi^{\star}_{MJ\Sigma}$
and $\Psi_{M'J'\Sigma'}$ stem from the velocity-state representations of the
baryon states $\left<P,J,\Sigma\right|$ and $\left|P',J',\Sigma'\right>$, 
respectively. These wave functions depend on the quark momenta
${\vec k}_i$ for which
$\sum_i{{\vec k}_i}=\vec 0$. They are related to the individual quark 
momenta by the Lorentz boost relations $p_i=B\left(v\right)k_i$.
The main challenge lies in the definition of a consistent
and reasonable momentum-space representation of the decay operator
$ \hat {\cal D}_{\alpha}$. Here, we adopt the PFSA and proceed in 
analogy 
to previous studies of the electroweak nucleon 
structure~\cite{Wagenbrunn:2000es,Glozman:2001zc,Boffi:2001zb}
but use a pseudovector coupling at the quark-pion vertex:
\begin{multline}
    \left<{p_{1}},{p_{2}},
    {p_{3}};\sigma_{1},\sigma_{2},\sigma_{3}\right|
    \hat {\cal D}_{\alpha}
    \left|{ p'_{1}},{ p'_{2}},
    { p'_{3}};\sigma'_{1},\sigma'_{2},\sigma'_{3}\right>\\
   =\sqrt{\frac{
M^3{M'}^3
}{
\left(\sum{\omega_i}\right)^3\left(\sum{\omega'_i}\right)^3
}
}
3i g_{qq\pi}
    \bar u\left({p_{1}},\sigma_{1}\right)
    \gamma^{5}\gamma^{\mu}{\vec \lambda^{F}}
    u\left({ p'_{1}},\sigma'_{1}\right)\\
    2p'^{0}_{2} \delta\left(\vec p_{2}-{\vec p}_{2}^{\,}{\!}'\right)
   2p'^{0}_{3}\delta\left(\vec p_{3}-{\vec p}_{3}^{\,}{\!}'\right)
   \delta_{\sigma_{2}\sigma'_{2}}
   \delta_{\sigma_{3}\sigma'_{3}}Q_{\pi,\mu}
  \label{eq:decayPFSA}.
  \end{multline}
The overall momentum conservation, $P'_{\mu}-P_{\mu}=Q_{\pi,\mu}$, 
together with the two spectator conditions define the relation between
all incoming and outgoing quark momenta. In particular, the momenta of 
the active quark are related by $\vec p_1-\vec p_{1}^{\,}{\!}'=\vec {\tilde Q}$, 
where $\vec {\tilde Q}$ is completely determined. Thus the
momentum transferred to the active quark is different from
the momentum transfer to the baryon as a whole. This is a consequence 
of translational invariance which thereby introduces effective
many-body contributions into the above definition of the 
spectator-model decay operator.
Furthermore, in eq.~(\ref{eq:decayPFSA}) there appears an overall
normalization factor 
\begin{equation}
{\cal N}=\sqrt{\frac{
M^3{M'}^3
}{
\left(\sum{\omega_i}\right)^3\left(\sum{\omega'_i}\right)^3
}
}
\label{eq:renorm}.
\end{equation}
Through the $\omega_i$ and the on-mass-shell condition of the quarks
it depends on the individual quark momenta. This choice of $\cal N$ is 
consistent with the one used in the definition of the electromagnetic 
and axial currents in the PFSA calculations of the nucleon electroweak 
form factors by the Graz-Pavia
collaboration~\cite{Wagenbrunn:2000es,Glozman:2001zc,Boffi:2001zb}.
It guarantees for the correct proton charge. However, it is not a 
unique choice. Any other normalization factor of the asymmetric form
\begin{equation}
{\cal N}\left(y\right)=
\left(\frac{M^3}{\left(\sum{\omega_i}\right)^3}\right)^y
\left(\frac{{M'}^3}{\left(\sum{\omega'_i}\right)^3}\right)^{1-y}
\label{eq:offsymm}
\end{equation}
would do the same. In order to study the effects of these further 
choices we investigate the dependence of the decay widths on the 
parameter range $0\le y \le 1$.

\vspace{0.6cm}
\begin{center}
{\bf \large Results}
\end{center}
\vspace{0.3cm}

\begin{figure}[hpt]
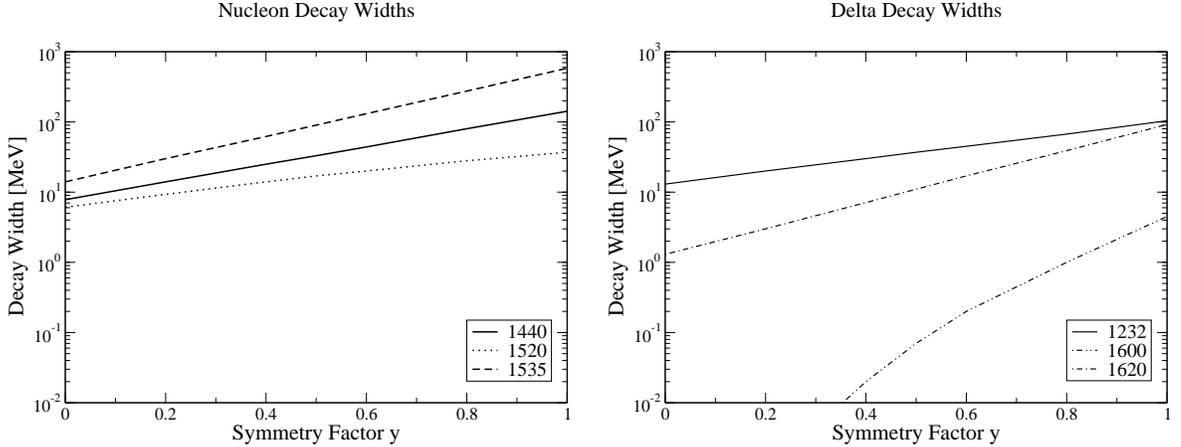

{
\includegraphics
[width=7.5cm]
{nucleon_off.eps}
\hspace{2mm}
\includegraphics
[width=7.5cm]
{delta_off.eps}
}
\caption{Dependence of some resonance decay widths on the choice of 
the normalization factor after eq.~(\ref{eq:offsymm}). 
}\label{fig:symmfac}
\end{figure}

The decay widths of the PFSA calculation with the decay operator
given in eq.~(\ref{eq:decayPFSA}), with the symmetric normalization 
factor, are shown in table~\ref{tab1} for the GBE and OGE CQMs.
It is immediately seen that only the $N^*_{1535}$ and
$N^*_{1710}$ predictions are
within the experimental range. All other decay widths are 
underestimated, some of them considerably. In this regard, it is
noteworthy that in the case of the  $N^*_{1535}$ the
$\Delta\pi$ branching ratio is exceptionally small ($<1\%$). Therefore 
we found it interesting to look at the results with a view to the 
measured $\Delta\pi$ branching ratios. In fact, one can observe a 
striking relation between these branching ratios and the sizes of the 
theoretical decay widths, expressed as 
percentage fractions of the experimental values in 
the last three columns of table~\ref{tab1}: The larger the $\Delta\pi$ 
branching ratio of a resonance, the bigger the underestimation of 
the (best-estimate) experimental value. This observation hints to a 
possible systematic problem in the description of mesonic decay widths 
within (relativistic) CQMs. It calls for a more complete treatment of 
baryon resonances with a more realistic coupling to decay channels. 
\renewcommand{\arraystretch}{1.5}
\begin{table*}[t!]
\caption{
PFSA predictions for $\pi$ decay widths of the relativistic GBE 
\protect\cite{Glozman:1998ag} and OGE \protect\cite{Theussl:2000sj} CQMs
in comparison to the Bethe-Salpeter results of the II CQM 
\protect\cite{Metsch:2004qk} and experimental data
\protect\cite{Eidelman:2004wy}.
In the last three columns the theoretical results are expressed as 
percentage fractions of the (best-estimate) experimental values in 
order to be compared to the measured $\Delta \pi$ branching
ratios. 
\label{tab1}
}
{\begin{tabular}{@{}crccccccc@{}}
\hline
Decays&Experiment&\multicolumn{3}{c}{
Rel. CQM
}&
$\Delta\pi$
&\multicolumn{3}{c}{
$\%$ of Exp. Width
}\\
{$\rightarrow N\pi $}&{[MeV]}& GBE & OGE & II & {\small branching ratio} & GBE & OGE & II\\
[0.25ex]
\hline
$N^{\star}_{1440}$
&
$\left(227\pm 18\right)_{-59}^{+70}$ &
$33$ &
$53$ &
$38$ &
$20-30\%$ & 
$14$ &
$24$ &
$17$\\ 
$N^{\star}_{1520}$
&
$\left(66\pm 6\right)_{-\phantom{0}5}^{+\phantom{0}9}$&
$17$ &
$16$ &
$38$ &
$15-25\%$ & 
$26$ &
$24$ &
$58$\\ 
$N^{\star}_{1535}$
&
$ \left(67\pm 15\right)_{-17}^{+28}$&
$90$ &
$119$ &
$33$ &
$<1\%$ & 
$134$ &
$178$ &
$49$\\ 
$N^{\star}_{1650}$
&
$\left(109\pm 26 \right)_{-\phantom{0}3}^{+36}$&
$29$ &
$41$ &
$\phantom{0}3$ &
$1-7\%$ & 
$27$ &
$38$ &
$\phantom{0}3$\\ 
$N^{\star}_{1675}$
&
$ \left(68\pm 8\right)_{-\phantom{0}4}^{+14}$&
$5.4$ &
$6.6$ &
$\phantom{0}4$ &
$50-60\%$ & 
$\phantom{0}8$ &
$10$ &
$\phantom{0}6$\\ 
$N^{\star}_{1700}$
&
$ \left(10\pm 5\right)_{-\phantom{0}3}^{+\phantom{0}3}$&
$0.8$ &
$1.2$ &
$0.1$ &
$>50\%$ & 
$\phantom{0}8$ &
$12$ &
$\phantom{0}1$\\ 
$N^{\star}_{1710}$
&
$\left(15\pm 5\right)_{-\phantom{0}5}^{+30}$&
$5.5$ &
$7.7$ &
$n/a$ &
$15-40\%$ & 
$37$ &
$51$ &
$n/a$\\ 
$\Delta_{1232}$
&
$\left(119\pm 1 \right)_{-\phantom{0}5}^{+\phantom{0}5}$&
$37$ &
$32$ &
$62$ &
$-$ & 
$31$ &
$27$ &
$52$\\ 
$\Delta_{1600}$
&
$\left(61\pm 26\right)_{-10}^{+26}$&
$0.07$ &
$1.8$ &
$n/a$ &
$40-70\%$ & 
$\approx 0$ &
$3$ &
$n/a$\\ 
$\Delta_{1620}$
&
$\left(38\pm 8\right)_{-\phantom{0}6}^{+\phantom{0}8}$&
$11$ &
$15$ &
$\phantom{0}4$ &
$30-60\%$ & 
$29$ &
$39$ &
$11$\\ 
$\Delta_{1700}$
&
$\left(45\pm 15\right)_{-10}^{+20}$&
$2.3$ &
$2.3$ &
$\phantom{0}2$ &
$30-60\%$ & 
$\phantom{0}5$ &
$\phantom{0}5$ &
$\phantom{0}4$\\
\hline
\end{tabular}}
\end{table*}
In fig.~\ref{fig:symmfac} we demonstrate the dependence of the PFSA 
predictions (for the case of the GBE CQM) on the possible asymmetric 
choice of the normalization factor $\cal N$ (see 
eq.~(\ref{eq:offsymm})). In the range $0\le y\le 1$ all decay widths 
grow rapidly with increasing $y$. In this way one could enhance the 
theoretical predictions considerably. However, if one wants neither 
one of the decay widths to exceed its experimental range, one is 
limited to a value of $y\le 0.5$. Any $y$ lower than 0.5 would lead
to decay widths much too small in most cases. Consequently, a
symmetric normalization factor as in eq.~(\ref{eq:decayPFSA}) seems
to be the preferred and most reasonable choice also in the context of 
hadronic decay widths.

\vspace{0.6cm}
\begin{center}
{\bf \large Summary}
\end{center}
\vspace{0.3cm}

We have presented a Poincar\'e-invariant description of strong baryon 
resonance decays in point form within relativistic CQMs.
Covariant predictions have been
given for $\pi$ decay widths. They are considerably different from 
previous nonrelativistic results or results with relativistic 
corrections included. The covariant results calculated with a 
spectator-model decay operator show a uniform trend.
In almost all cases the corresponding theoretical predictions
underestimate the experimental data considerably. This is true in the
framework of Poincar\'e-invariant quantum mechanics (here in point form)
as well as in the Bethe-Salpeter approach \cite{Metsch:2004qk}.
Indications have been given that for a particular resonance the size
of the underestimation is related to the magnitude of the $\Delta\pi$ 
branching ratio. This hints to a systematic defect in the description 
of the decay widths.

The investigation of different possible choices for a normalization 
factor in the spectator-model decay operator has led to the 
suggestion that the symmetric choice is the most natural one. It is 
also consistent with the same (symmetric) choice that had been adopted
before for the spectator-model current in the study of the electroweak
nucleon form factors.

\vspace{0.3cm}
\begin{center}
{\small This work was supported by the Austrian Science Fund (Project 
P16945).
T.M. would like to thank the 
INFN and the Physics Department of the University of Padova for their 
hospitality, and MIUR-PRIN for financial support}
\end{center}
\end{document}